\documentclass[onecolumn,showpacs,preprintnumbers,amsmath,amssymb]{revtex4}
\usepackage{graphicx}
\usepackage{dcolumn}
\usepackage{bm}

\begin{document}

\title{Quantum teleportation via a $W$ state} 

\author{Jaewoo Joo} \email{j-w-joo@hanmail.net}

\affiliation{Department of Physics, Sogang University, Seoul 121-742, Korea}

\author{Young-Jai Park} \email{yjpark@ccs.sogang.ac.kr}
\affiliation{Department of Physics, Sogang University, Seoul 121-742, Korea}

\author{Sangchul Oh} \email{scoh@kias.re.kr}

\affiliation{School of Computational Sciences, Korea Institute for Advanced Study,
Seoul 130-722, Korea}

\author{Jaewan Kim} \email{jaewan@kias.re.kr}

\affiliation{School of Computational Sciences, Korea Institute for Advanced Study,
Seoul 130-722, Korea}

\date{\today}

\begin{abstract}
We investigate two schemes of the quantum teleportation with a $W$ state, which belongs to 
a different class from a Greenberger-Horne-Zeilinger class. In the first scheme, the $W$ state is shared by three 
parties one of whom, called a sender, performs a Bell measurement. It is shown that 
quantum information of an unknown state is split between two parties and recovered with 
a certain probability. In the second scheme, a sender takes two particles of the $W$ 
state and performs positive operator valued measurements. For two schemes,
we calculate the success probability and the average fidelity. We show that
the average fidelity of the second scheme cannot exceed that of the first one.
\end{abstract}
\pacs{03.65.Bz, 03.67.-a, 03.67.Hk}

\maketitle
\section{Introduction}
Entanglement is considered to be a fundamental resource of 
quantum information processing such as quantum cryptography 
\cite{Bennett84}, quantum teleportation \cite{Bennett93},
quantum computation \cite{Neilsen00} and so on. 
Since Bennett {\em et al.}'s seminal work \cite{Bennett93},
there have been intensive works on the field of quantum
teleportation in theories and experiments \cite{Braunstein, Stenholm, Lee02}.
Entangled states, which are called quantum channels, make it possible to send an unknown state
to a long distance. 

Quantum teleportation via bipartite pure entangled states can be classified
into two types; standard teleportation (ST) and conclusive teleportation (CT)
\cite{Mor99, Bandyopadhyay00, Son01}.
Maximally entangled states, {\em i.e.} Bell states, are used as quantum
channels in standard quantum teleportation. On the other hand conclusive
teleportation utilizes any pure entangled states. Due to the fact that
non-maximally pure entangled states are used as quantum channels,
the random outcome of sender's measurement determines whether an
unknown state is teleported surely or not.                

Entanglement in three qubits is more complicated than that in two qubits. 
Recently, the entanglement of three qubits~\cite{Dur99, Acin01} 
was classified by separable, bi-separable, $W$, and Greenberger-Horne-Zeilinger (GHZ) 
states~\cite{Coffman00, Dur00, GHZ90}. The GHZ class state cannot be transformed
to the $W$ class by the local operation and classical communication. 
Although many suggestion has utilized the GHZ state 
in quantum teleportation~\cite{Karlsson98, Hillery99, Shi00} 
and some proposals have suggested an implementation of the $W$ state 
\cite{Zeilinger97, Guo02, Wang, Gorbachev03},
the teleportation using the $W$ class state 
has been studied in a few ways~\cite{Gorba02}.

In this paper, we study two schemes of teleportation via a $W$ state 
which are classified by two types of measurements. By calculating the total 
success probability and the average fidelity for each teleportation scheme,
we can explore the entanglement properties of $W$ state. Because of the property 
of the $W$ state \cite{Joo02}, it is shown to be not always successful in recovering 
of the unknown state.

In the first scheme, Alice teleports an unknown state to Charlie with help of Bob. 
She performs the Bell measurement (BM) on the unknown state and her part of
a $W$ state. This scheme makes it possible to split and to recover quantum information of 
the unknown state in the teleportation. After Alice's BM, the quantum information 
of the unknown state is split between Bob and Charlie. Depending on Bob's 
measurement on his part of the $W$ state, Charlie recovers the unknown state 
with a certain probability of success.

In the second scheme, only Alice and Charlie participate in teleportation. 
She and he take two particles and one particle of the $W$ state, respectively. 
Alice prepares a basis set of an unbalanced $W$ state the positive operator valued
measurement (POVM) \cite{Peres93} on her three particles including the unknown
state in teleportation. 
This type of the measurement is called an asymmetic POVM because 
the measurement basis is the $W$-type with unequal weights. 
The asymmetric POVM is fully based on an original 
CT \cite{Mor99, Son01} where the measurement set is built from 
an orthogonality between a measurement basis and a channel.
From calculation of success probability and average fidelity, we obtain the maximum values
of the protocol, which can not exceed these of the previous scheme.

The paper is organized as follows. In Sec.~\ref{sec:GHZ}, we briefly review quantum 
teleportation via the GHZ state. In Sec.~\ref{sec:W_state}, we present two schemes of quantum 
teleportation with the $W$ state. In Sec.~\ref{sec:fidelity}, the average fidelity of 
two schemes is calculated. Finally, in Sec.~\ref{sec:remarks} we make some remarks on our results.

\section{Quantum teleportation via GHZ states}
\label{sec:GHZ}

We start with a brief review of quantum teleportation with a GHZ state in Ref.~\cite{Karlsson98,Hillery99}. 
It is similar to the ST with the Bell state, but three parties participate in
quantum teleportation. Especially, Hillery, Bu$\check{z}$ek, and Berthiaume emphasized quantum teleportation with the GHZ state as splitting and recovering of quantum information as a quantum secret sharing~\cite{Hillery99}.
 
Suppose Alice wishes to teleport an unknown qubit to Charlie with assistance of Bob.
Alice begins with the unknown state 
\begin{equation}
\label{eq:uk}
|\phi \rangle_{U} = \alpha |z_- \rangle_{U}+\beta |z_+ \rangle_{U},
\end{equation}
where $|\alpha|^2 + |\beta|^2 = 1$ and $z_{\pm}$ are the outcomes ($+$ or $-$) 
in measuring $\hat{\sigma}_z$. Alice, Bob, and Charlie share the GHZ state 
given by
\begin{eqnarray}
\label{eq:ghz}
|{\rm GHZ} \rangle \equiv \frac{1}{\sqrt{2}} ( |z_- z_- z_- \rangle + |z_+ z_+ z_+ \rangle ).
\end{eqnarray}
Alice combines the unknown state $|\phi\rangle_{U}$ with her GHZ particle
and performs the BM on her two particles. Before Alice's BM, 
the total state $|\Psi_{\rm GHZ}^{\rm tot} \rangle = |\phi \rangle_{U}
\otimes|{\rm GHZ}\rangle_{ABC}$ can be written in terms of Bell basis of her part

\begin{eqnarray}
\label{eq:GHZ_tot}
|\Psi_{\rm GHZ}^{\rm tot} \rangle = \frac{1}{2} \bigl[
& |\Psi_{+}\rangle_{UA} 
(\alpha | z_- z_- \rangle + \beta |z_+ z_+ \rangle) + |\Psi_{-}\rangle_{UA} 
(\alpha | z_- z_- \rangle - \beta |z_+ z_+ \rangle) & \nonumber \\
+&  |\Phi_{+}\rangle_{UA} 
(\beta | z_- z_- \rangle + \alpha |z_+ z_+ \rangle) + |\Phi_{-}\rangle_{UA} 
(\beta | z_- z_- \rangle - \alpha |z_+ z_+ \rangle) & \bigr],
\end{eqnarray}
where Bell basis states are given by 
\begin{eqnarray}
\label{eq:Bellstate}
|\Psi_{\pm} \rangle_{UA} & = & \frac{1}{\sqrt{2}}
(|z_- z_- \rangle_{UA}\pm |z_+ z_+ \rangle_{UA}), \nonumber \\
|\Phi_{\pm}\rangle_{UA}  & = &  \frac{1}{\sqrt{2}}
(|z_- z_+ \rangle_{UA}\pm |z_+ z_- \rangle_{UA}).
\end{eqnarray}
After Alice's BM, the total state is projected into one of four states 
in Eq.~(\ref{eq:GHZ_tot}). It is interesting that quantum information of an unknown state 
is split between Bob and Charlie at this moment and recovered in Charlie's place 
by Bob's measurement and Charlie's unitary operation.

Bob measures his GHZ particle in $x$-direction and announces his outcome.
With the information of two classical bits from Alice and one bit from Bob, Charlie can 
make a suitable unitary operation on his qubit to recover the original unknown state 
as follows
\begin{eqnarray}
\label{eq:ghz_arrow}
\begin{array}{lll}
 |\Psi_{+}\rangle_{UA}|x_+ \rangle_{B} \rightarrow  & I, 
&|\Psi_{+}\rangle_{UA}|x_- \rangle_{B} \rightarrow \hat{\sigma}_{z}, \\[6pt]
 |\Phi_{+}\rangle_{UA}|x_+ \rangle_{B} \rightarrow & \hat{\sigma}_{x}, 
&|\Phi_{+}\rangle_{UA}|x_- \rangle_{B} \rightarrow \hat{\sigma}_{x}\hat{\sigma}_{z}, \\[6pt]
 |\Psi_{-}\rangle_{UA}|x_+ \rangle_{B} \rightarrow & \hat{\sigma}_{z}, 
&|\Psi_{-}\rangle_{UA}|x_- \rangle_{B} \rightarrow  I,  \\[6pt]
 |\Phi_{-}\rangle_{UA}|x_+ \rangle_{B} \rightarrow & \hat{\sigma}_{x} \hat{\sigma}_{z}, 
 \hspace{0.75cm}
&|\Phi_{-}\rangle_{UA}|x_- \rangle_{B} \rightarrow \hat{\sigma}_{x} .
\end{array}
\end{eqnarray}
As a result, in the BM of Alice's part, 
the teleportation using the GHZ state is always in success.

On the other hand, suppose Alice and Charlie wish to make teleportation via a GHZ state without Bob.
Alice takes two particles of the GHZ state $|{\rm GHZ}\rangle_{AA'C}$ and Charlie the other one.
Teleportation can be implemented if Alice changes her set of measurement from BMs to 
the GHZ-type measurement as follows,
\begin{eqnarray}
|\Psi_{\pm}^{\rm GHZ} \rangle_{UAA'}  
&=& \frac{1}{\sqrt{2}} (|z_- z_- z_-  \rangle \pm |z_+ z_+ z_+  \rangle ), \nonumber \\
|\Phi_{\pm}^{\rm GHZ} \rangle_{UAA'}  
&=& \frac{1}{\sqrt{2}} (|z_- z_+ z_+  \rangle \pm |z_+ z_- z_-  \rangle ).
\end{eqnarray}
If we use a notation $|z_- \rangle_{A}$ ($|z_+ \rangle_{A}$) in the place of 
$|z_- z_-  \rangle_{AA'}$ ($|z_+ z_+  \rangle_{AA'}$),
the protocol becomes exactly same as ST. Therefore, we are able to perform a perfect teleportation via the GHZ state. 

Furthermore, what about teleportation of a partially entangled GHZ state? In fact, Bandyopadhyay's work implies this problem in Ref.~\cite{Bandyopadhyay00}. He introduced an assisted qubit, which has an equivalent coefficient to a partially entangled channel and Alice measures the unknown qubit with the assisted qubit and her channel qubit. The scheme shows a method of the teleportation via a partially entangled GHZ state.

\section{Quantum teleportation via $W$ states}
\label{sec:W_state}

\subsection{Bell measurement} 
\label{sec:BellW}
A similar idea of quantum teleportation applies to the following $W$ state
\begin{eqnarray}
  \label{eq:ws}
  | W \rangle = \frac{1}{\sqrt{3}} 
              \left( |z_+ z_- z_- \rangle + |z_-z_+z_-\rangle + |z_-z_-z_+\rangle \right).
\end{eqnarray}
Alice also begins with an unknown state in Eq.~(\ref{eq:uk}). 
Alice, Bob, and Charlie share the $W$ state.
Alice combines an unknown state with her $W$ qubit and performs BM on these qubits.
The basis vectors of Bell state operators in $z$-direction 
are represented by Eq.~(\ref{eq:Bellstate}). Before Alice's measurement, the total state 
$|\Psi_{ W}^{\rm tot} \rangle = |\phi \rangle_{U} \otimes |W \rangle_{ABC}$ 
can be expressed in terms of the Bell basis as follows 
\begin{widetext}
\begin{eqnarray}
\label{eq:tot_w}
|\Psi_{ W}^{\rm tot} \rangle 
= \frac{1}{\sqrt{6}} \Bigl[ 
&|\Psi_{+}\rangle_{UA} \{ |z_-  \rangle_{B} ( \beta |z_- \rangle_{C} + 
\alpha |z_+ \rangle_{C} ) + |z_+ \rangle_{B} (\alpha | z_- \rangle_{C}) \}& \nonumber \\
-&|\Psi_{-} \rangle_{UA} \{|z_-  \rangle_{B}( \beta |z_- \rangle_{C} - \alpha |z_+ \rangle_{C}) - |z_+ \rangle_{B} (\alpha  |z_- \rangle_{C} ) \}& \nonumber \\[6pt]
+&|\Phi_{+} \rangle_{UA} \{|z_-  \rangle_{B} ( \alpha |z_- \rangle_{C} + \beta |z_+ \rangle_{C} ) + |z_+ \rangle_{B} (\beta | z_- \rangle_{C} ) \}& \nonumber \\
+&|\Phi_{-} \rangle_{UA} \{|z_-  \rangle_{B} ( \alpha |z_- \rangle_{C} - \beta |z_+ \rangle_{C} ) - |z_+ \rangle_{B} ( \beta | z_- \rangle_{C} ) \}&\!\Bigr ]. 
\end{eqnarray}
\end{widetext}
After Alice's measurement, the total state is collapsed into one of the four states in Eq.~(\ref{eq:tot_w}).
Unlike the ST with maximally entangled Bell state or with a GHZ state, 
probabilities of four outcomes depend on the unknown state. The probability $p_{\Psi_{+}}$ that Alice's output 
state is $|\Psi_{+}\rangle$ is given by
\begin{eqnarray}
p_{\Psi_{+}} = |_{UA}\langle \Psi_{+} | \Psi_{ W}^{\rm tot} \rangle_{UABC} |^2 =
\frac{1+|\alpha|^2}{6}.
\end{eqnarray}
Also, probabilities of Alice's outcomes as 
$|\Psi_{-}\rangle$, $|\Phi_{+}\rangle$ and 
$|\Phi_{-}\rangle$ are $(1+|\alpha|^2)/6$, $(1+|\beta|^2)/6$, and $(1+|\beta|^2)/6$, 
respectively.  

Bob's measurement outcome dicides success or failure of teleportation. 
If Bob's outcome is $z_+$, the teleportatioin fails. Otherwise, Charlie can do the following 
operation on his $W$ qubit to get the unknown state according to Alice's information.
\begin{eqnarray}
\label{eq:wrelation}
\begin{array}{ll}
 |\Psi_{+}\rangle_{UA}|z_- \rangle_{B} \rightarrow \hat{\sigma}_{x},\quad
&|\Psi_{-}\rangle_{UA}|z_- \rangle_{B} \rightarrow \hat{\sigma}_{x} \hat{\sigma}_{z},\\[6pt]
 |\Phi_{+}\rangle_{UA}|z_- \rangle_{B} \rightarrow I,
&|\Phi_{-}\rangle_{UA}|z_- \rangle_{B} \rightarrow  \hat{\sigma}_{z}.
\end{array}
\end{eqnarray}

Recently, it has been proposed as a scheme of teleportation via a $W$ state 
\cite{Shi02}. In fact, they have suggested two ways of teleportation which have 
different probabilities, but we have pointed out that two schemes give the same probability
as 2/3, which does not depend on coefficients $\alpha$ and $\beta$ of the unknown state \cite{Joocom}.

When Alice has an outcome state $|\Psi_{+} \rangle$, the probability that Bob measures 
an outcome $z_-$ is $1/(1+|\alpha|^2)$. When we define a probability $p_{1}$ ($p_{2}$, 
$p_{3}$, $p_{4}$) as $|_{B} \langle z_- |_{UA}\langle \Psi_{+} |\Psi _{W}^{\rm tot} \rangle_{UABC}|^2$ (with the substitutions of $\Psi_{+}$ by  $\Psi_{-}$,  $\Phi_{+}$,
 $\Phi_{-}$), the success probability of the scheme is given by
\begin{eqnarray}
\label{eq:con}
P_{\rm bell}^{\rm suc} &=& \sum_{i=1}^{4} p_{i} = 2 \cdot \frac{1}{1+|\alpha|^2} \cdot \frac{1+|\alpha|^2}{6} 
+ 2 \cdot \frac{1}{1+|\beta|^2} \cdot \frac{1+|\beta|^2}{6} = \frac{2}{3}.
\end{eqnarray}

On the other hand, probabilities of failed teleportation are given by
\begin{eqnarray}
\label{eq:incon}
p_{5} &=& |_{B} \langle z_+ |_{UA} \langle \Psi_{+} | \Psi _{ W}^{\rm tot} \rangle_{UABC} |^2 
       = \frac{|\alpha|^2}{6},\nonumber \\
p_{6} &=& |_{B} \langle z_+ |_{UA} \langle \Psi_{-} | \Psi _{ W}^{\rm tot} \rangle_{UABC}|^2 
       = \frac{|\alpha|^2}{6}, \nonumber \\
p_{7} &=& |_{B} \langle z_+ |_{UA} \langle \Phi_{+} | \Psi _{W}^{\rm tot} \rangle_{UABC}|^2 
       = \frac{|\beta|^2}{6}, \nonumber \\
p_{8} &=& |_{B} \langle z_+ |_{UA} \langle \Phi_{-} | \Psi _{W}^{\rm tot} \rangle_{UABC}|^2 
       = \frac{|\beta|^2}{6},
\end{eqnarray}
where the total probability of failure is $P_{\rm bell}^{\rm fail} = \sum_{i=5}^{8} p_{i} = 1/3$.

Even though the scheme does not perform POVMs \cite{Son01}, 
it can also be called a kind of CT by $W$ states
because a conclusive event occurs.
The total success probability of teleportation is 2/3 in the scheme 
because the probability of $z_-$ outcome of Bob is 2/3 \cite{Joocom}.
In fact, it is a natural conclusion that Bob's outcome decides 
whether the quantum channel between Alice and Charlie is entangled or not.

\subsection { Asymmetric $W$-type measurements} 
\label{sec:chanW}

Let us suppose only Alice and Charlie participate in quantum teleportation via 
the $W$ state. Alice keeps two qubits of the $W$ state and Charlie the other one.
Although ST gives us a perfect teleportation via an EPR pair,
we would like to compare a feature of teleportation by BM with that by POVM. 
Alice performs three-qubit POVMs on one qubit of the unknown state and 
her two qubits of the $W$ state. Notice that the protocol of CT of 
a $d$-dimensional state with POVMs is developed in a way of making a conclusive 
event~\cite{Son01}. Similarly, we are able to perform a way of the POVMs without any loss of generality.

The total state of the unknown state and the $W$ state can be expressed in terms of 
the bases as follows
\begin{eqnarray}
|\Psi_{W}^{\rm tot} \rangle_{UAA'C} 
&=& \frac{1}{2} \{~|\psi^{'}_{1} \rangle (\alpha |z_- \rangle + \beta |z_+ \rangle) 
+|\psi^{'}_{2} \rangle (\alpha |z_- \rangle- \beta |z_+ \rangle) \nonumber \\
&&~+|\psi^{'}_{3} \rangle (\beta |z_- \rangle + \alpha |z_+ \rangle) 
+|\psi^{'}_{4} \rangle (\beta |z_- \rangle - \alpha |z_+ \rangle) \},
\end{eqnarray}
where the basis vectors are represented by 
\begin{eqnarray}
|\psi^{'}_{1} \rangle_{UAA'}  
= \frac{1}{\sqrt{3}} ( |z_- z_- z_+\rangle + |z_- z_+ z_-\rangle +  |z_+ z_- z_-\rangle),  \nonumber \\
|\psi^{'}_{2} \rangle_{UAA'}  
= \frac{1}{\sqrt{3}} ( |z_- z_- z_+\rangle + |z_- z_+ z_-\rangle -  |z_+ z_- z_-\rangle), \nonumber \\
|\psi^{'}_{3} \rangle_{UAA'} 
= \frac{1}{\sqrt{3}} ( |z_+ z_- z_+\rangle + |z_+ z_+ z_-\rangle +  |z_- z_- z_-\rangle), \nonumber \\
|\psi^{'}_{4} \rangle_{UAA'} 
= \frac{1}{\sqrt{3}} ( |z_+ z_- z_+\rangle + |z_+ z_+ z_-\rangle -  |z_- z_- z_-\rangle).
\end{eqnarray}

For Alice's POVMs, we consider unnormalized vectors $|\tilde{\psi^{'}}_{i}\rangle$
to satisfy a condition of $\langle \tilde{\psi^{'}}_{i} |\psi^{'}_{j}\rangle = \delta_{ij}$.
Generally, there are three parameters in each $|\tilde{\psi^{'}}_{i} \rangle$ but the condition
reduces to one parameter such as $a$ or $a'$. The basis vectors of the asymmetric POVM
are represented 
\begin{widetext}
\begin{eqnarray}
|\tilde{\psi^{'}}_{1} \rangle_{UAA'}  
&=&  a |z_- z_- z_+\rangle + (\frac{\sqrt{3}}{2} - a)  |z_- z_+ z_-\rangle
+ \frac{\sqrt{3}}{2}  |z_+ z_- z_-\rangle, \nonumber \\
|\tilde{\psi^{'}}_{2} \rangle_{UAA'}  
&=&  a |z_- z_- z_+\rangle  + (\frac{\sqrt{3}}{2} - a) |z_- z_+ z_-\rangle 
- \frac{\sqrt{3}}{2}  |z_+ z_- z_-\rangle, \nonumber \\
|\tilde{\psi^{'}}_{3} \rangle_{UAA'}  
&=&  a'|z_+ z_- z_+ \rangle\ + (\frac{\sqrt{3}}{2} - a') |z_+ z_+ z_-\rangle 
   + \frac{\sqrt{3}}{2}  |z_- z_- z_-  \rangle, \nonumber \\
|\tilde{\psi^{'}}_{4} \rangle_{UAA'}  
&=&  a'|z_+ z_- z_+\rangle + (\frac{\sqrt{3}}{2} - a')  |z_+ z_+ z_-\rangle 
- \frac{\sqrt{3}}{2}|z_- z_- z_-\rangle).
\end{eqnarray}
\end{widetext}
The proper POVM set $ \{ \hat{M}_{i} \}$ is defined by
\begin{eqnarray}
\hat{M}_{i} &=& \lambda_{\rm asym}  | \tilde{\psi^{'}}_{i} \rangle \langle \tilde{\psi^{'}}_{i} |,
\nonumber \\
\hat{M}_{5} &=& \openone_{8 \times 8} - \sum_{i=1}^{4} \hat{M}_{i},~~~
 (\langle \tilde{\psi^{'}}_{i} |\psi^{'}_{j}\rangle = \delta_{ij}),
\end{eqnarray}
where the real parameter $ \lambda_{\rm asym}  \geq 0  $ and $ i = 1, 2, 3, 4$.
The set of this POVM also satisfies positivity and completeness 
for any quantum state $|\psi^{'} \rangle$ as follows
\begin{eqnarray}
\label{eq:lam1}
\sum_{i=1}^{4} \lambda_{\rm asym}  | \langle \tilde{\psi^{'}}_{i} | \psi \rangle |^2 \le 1~~\rm{for} ~\forall |\psi \rangle.
\end{eqnarray}

For simplicity, let us assume that $a$ is real and equal to $a'$.
Since $\hat{M}_{5}$ should be a positive operator, $a$ and $\lambda_{\rm asym}$
should satisfy one of the following conditions as Fig. 1.
\begin{eqnarray}
\label{eq:parameter}
({\rm i}) & ~ 0<a \le \frac{\sqrt{3}}{2}  \hspace{0.35cm} & 
0 < \lambda_{\rm asym}  \le \frac{2}{3} \nonumber \\
({\rm ii}) &~ \frac{\sqrt{3}}{2} < a   \hspace{0.35cm} & 
0 < \lambda_{\rm asym}  \le \frac{1}{4 a^2 - 2  \sqrt{3} a + \frac{3}{2}}. 
\end{eqnarray}

\begin{figure}[ht]
\label{fig:wpovm12}
\centering
%\vspace{0.1in}
\includegraphics[width= 3.1in]{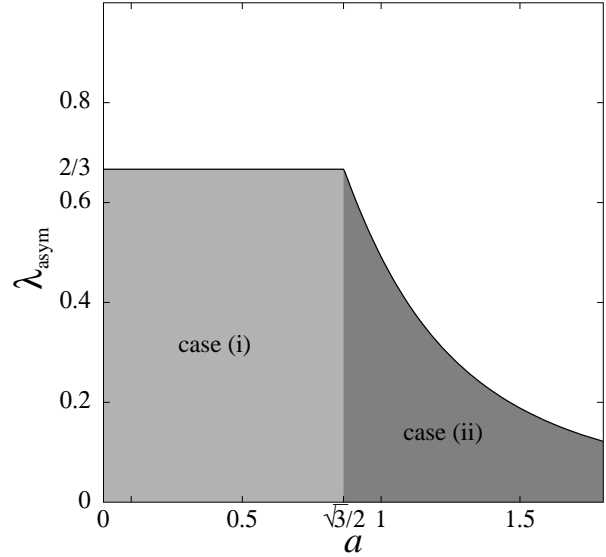}
\caption{Parameter range of an asymmetric POVM (Eq. \ref{eq:parameter})}
\end{figure}

We can easily calculate the probability $p_{i}$ of each conclusive case and total success probability $P_{\rm asym}^{\rm suc}$
\begin{eqnarray}
p_{i} &=&  \langle \Psi_{W}^{\rm tot}| \hat{M}_{i} |\Psi_{W}^{\rm tot} \rangle 
= \frac{\lambda_{\rm asym} }{4} \\
P_{\rm asym}^{\rm suc} &=& \sum_{i=1}^4 p_{i} = \lambda_{\rm asym}  \le \frac{2}{3},
\end{eqnarray}
with total failure probability $P_{\rm asym}^{\rm fail} = 1 -  P_{\rm asym}^{\rm suc}$.
Since $\lambda_{\rm asym} $ is at most 2/3 in Fig. 1, the optimal success teleportation with
an asymmetric POVM is 2/3. 

As a result of two cases, although we optimize a parameter in Alice's part, 
total success probabilities of each scheme in quantum teleportation
has a limit value of  2/3. Therefore, it can be characterized as a property 
of a $W$ state when it is used in quantum teleportation.
 
\section {Average fidelity of Teleportation with $W$ states}
\label{sec:fidelity}
In any protocol of teleportation, it is necessary to average the fidelity over all 
possible input states. The average fidelity reflects how much information is transferred 
to the teleported state. The teleported pure state of the density operator $\hat{\rho}_i$ 
relies on the outcome $i$. The average fidelity is defined as
\begin{eqnarray}
\bar{\cal F} \equiv \frac{1}{V} \int d{\vec \Omega} \sum_i
p_i ({\vec \Omega}) f_i ({\vec \Omega}),
\end{eqnarray}
where $f_i ({\vec \Omega}) = \langle \phi({\vec \Omega}) |
\hat{\rho}_{i} | \phi({\vec \Omega}) \rangle$ and an unknown pure
state $|\phi({\vec \Omega})\rangle$ is parameterized by a real vector
${\vec \Omega}$ in the parameter space of volume $V$
\cite{Son00}.
In the CT, 
we separate the average fidelity into two parts for easy calculation as follows
\begin{eqnarray}
\label{eq:fedelity1}
\bar{\cal F}&=& \bar{\cal F}_{\rm con} + \bar{\cal F}_{\rm inc}
= \frac{1}{V}\int d{\vec \Omega} \Big[\sum^{m}_{i=1}
  +\sum^{n+m}_{i=m+1} \Big] p_{i} ({\vec \Omega})
  f_{i} ({\vec \Omega}), 
\end{eqnarray} 
where $m$ and $n$ are the total numbers of conclusive events 
and inconclusive events.

To understand protocols clearly, 
we make a comparison of average fidelity in each case of quantum teleportation 
of the BM (III A) and the asymmetric POVM (III B).
Although Eq.~(\ref{eq:fedelity1}) contains integration of ${\vec \Omega}$,
probability $p_{i}({\vec \Omega})$ is independent of ${\vec \Omega}$ and
$ \frac{1}{V}\int d{\vec \Omega}  f_{i} ({\vec \Omega})$ is 1 for conclusive event 
and 1/2 for inconclusive event.

For the case of the teleportation by the BM, 
the success probability $p_{i}$ ($i = 1, 2, 3, 4$) in Eq.~(\ref{eq:con}) is 1/6 and
the sum of the failure probability $p_{i}$ ($i = 5, 6, 7, 8$) in Eq.~(\ref{eq:incon}) 1/3.
Then, the average fidelity in the BM is
\begin{eqnarray}
\label{eq:fidelity}
\bar{\cal F}_{\rm Bell}  
&=&  \sum^{4}_{i =1} \frac{1}{6} \cdot 1 + \sum^{8}_{i =5}  p_{i} \cdot \frac{1}{2} 
 = \frac{5}{6}. 
\end{eqnarray} 

In the teleportation by an asymmetric POVM, the success probability 
of each conclusive event is $ \lambda_{\rm asym} /4$ with fidelity 1.
The average fidelity is
\begin{eqnarray}
\bar{\cal F}_{\rm asym} = \sum^{4}_{i=1} \frac{\lambda_{\rm asym} }{4} \cdot 1
+ (1 - \lambda_{\rm asym} ) \cdot \frac{1}{2}  
= \frac{1}{2}+\frac{1}{2} \lambda_{\rm asym} \le \frac{5}{6}.
\end{eqnarray} 
Since $\lambda_{\rm asym} $ is smaller than 2/3, the 
maximal average fidelity is also 5/6 as a limit of the BM.
Therefore, the method of the BM provides the maximum value of average fidelity as well as that of success probability between two types of teleportations via a W state.
 
\section {Remarks}
\label{sec:remarks}

We have shown two schemes of quantum teleportation of an unknown qubit via a 
$W$ state taking into account two kinds of Alice's measurements.
In the first protocol, after Alice performs Bell measurements on the unknown state and
her channel state, the information of the unknown state is split between Bob and Charlie.
But recovering of the unknown state is not always successful unlike the quantum teleportation 
with a GHZ state. The second scheme of quantum teleportation is based on POVMs in Alice's 
part. We have also calculated the maximal success probability and the optimal average fidelity
for each scheme. The maximum values of success probability and average fidelity are 2/3 and 5/6. Since average fidelity 5/6 is greater than 2/3, the $W$ state can be used as a channel in quantum teleportation. Therefore, $W$ states are not particularly better for quantum teleportation than GHZ states.

Even though we can obtain same values of the success probability and the fidelity for each scheme,
we would like to comment that the amount of classical communication is different in each case.
The teleportation protocol via the $W$ state by the BM needs three classical bits
to achieve a successful teleportation, but the schemes by asymmetric POVM require three bits with consideration of a failure notice to a receiver.

Finally, we note a scheme of teleportation using a $W$ state of different amplitudes as follows
\begin{eqnarray}
|\Psi_{ W}\rangle =  a |z_- z_- z_+\rangle + b |z_- z_+ z_- \rangle + c |z_+ z_- z_-\rangle,
\end{eqnarray}
where $|a|^2+|b|^2+|c|^2=1$. 
Following the schemes shown in Sec.~\ref{sec:BellW}, we can easily guess a protocol of 
the BM from $|\Psi_{W}^{\rm tot}\rangle = 
|\phi\rangle_{U}\otimes |\Psi_{W}\rangle_{ABC}$. If Bob's measurement results in 
$|z_-\rangle$, Charlie also obtains a teleported state with  a certain probability. 
On the other hand, if we consider quantum teleportation with the POVM,
even in a simple case of $|a| = |c|$ and $|a| \ne |b|$, 
it seems to be difficult to find a POVM set.

\acknowledgments 
We thank J. Lee, W. Son, V. N. Gorbachev, and the KIAS Quantum Information Group
for useful discussion. This work of two of us (J. J. and Y.-J. P.) was supported 
by the Korea Research Foundation, Grant No. KRF-2002-042-C00010. 
J. K. was supported by Korea Research Foundation Grant No. KRF-2002-070-C00029.

\end{document}